\documentclass{article}
\usepackage{spconf,amsmath,epsfig}
\usepackage{color}
\usepackage[utf8]{inputenc}
\usepackage{subfigure}
\newcommand{\msf}{_\mathsf}

\newcommand{\my}{\mathbf{Y}}
\newcommand{\mx}{\mathbf{X}}
\newcommand{\mw}{\mathbf{W}}
\newcommand{\mk}{\mathbf{K}}

\newcommand{\mn}{\mathbf{N}}

\newcommand{\bt}[1]{\mbox{$\bf #1$}}

\begin{document}\sloppy

\def\x{{\mathbf x}}
\def\L{{\cal L}}

\title{On the Information Leakage of Camera Fingerprint Estimates}
\twoauthors{Samuel Fern\'andez-Mendui\~na}
{\textit{EE Department} \\
\textit{Imperial College}\\
London, UK \\
\tt{sf219@ic.ac.uk}}
{Fernando P\'erez-Gonz\'alez}
{\textit{Signal Processing in Communications Group} \\
\textit{Atlanttic Research Center}\\Vigo, Spain \\
\tt{fperez@gts.uvigo.es}}
%
%
%
%
\maketitle

\begin{abstract}
Camera fingerprints based on sensor PhotoResponse Non-Uniformity (PRNU) have gained broad popularity in forensic applications due to their ability to univocally identify the camera that captured a certain image. This fingerprint of a given sensor is extracted through some estimation method that requires a few images known to be taken with such sensor. In this paper, we show that the fingerprints extracted in this way leak a considerable amount of information from those images used in the estimation, thus constituting a potential threat to privacy. We propose to quantify the leakage via two measures: one based on the Mutual Information, and another based on the output of a membership inference test. Experiments with practical fingerprint estimators on a real-world image dataset confirm the validity of our measures and highlight the seriousness of the leakage and the importance of implementing techniques to mitigate it. Some of these techniques are presented and briefly discussed.    
\end{abstract}
\section{Introduction}
\label{sec:intro}

The PhotoResponse Non-Uniformity (PRNU) is a multiplicative spatial pattern that is present in every picture taken with a CCD/CMOS imaging device and acts as a unique fingerprint for the sensor itself \cite{Chen08}.  The PRNU is due to manufacturing imperfections that cause sensor elements to have minute area differences and thus capture different amounts of energy even under a perfectly uniform light field. The uniqueness of the PRNU has already led to a number of applications in multimedia forensics, both to solve camera identification/attribution problems and to detect inconsistencies that reflect intentional manipulations~\cite{KH17}.

Since the PRNU is a very weak signal, its extraction requires the availability of a number (often dozens) of images known to be taken with the camera under analysis. Although several extraction algorithms (both model- and data-driven) exist \cite{Chen08}, \cite{Cozzolino19}, all of them perform some sort of averaging across the residuals obtained by denoising the available images. The most prevalent method~\cite{Chen08} performs a further normalization to take into account the multiplicative nature of the PRNU. 

Unfortunately, the ease with which the PRNU can be extracted and the existence of relatively good theoretical models that explain its contribution lead to attacks that are similar in intention to {\em digital forgery attacks} in cryptography: the so-called {\em PRNU copy attack} plants the fingerprint from a desired camera in an image taken by a different device with the purpose of incriminating someone or merely undermining the credibility of PRNU-based forensics.  

While the PRNU copy attack can be considered a threat to {\em trust}, in this paper we identify risks to {\em privacy} by showing that there is substantial information leakage into the PRNU from the images used for its estimation. The existence of this leakage has been already indirectly exploited in the so-called {\em triangle test} \cite{triangle_test}, which is a countermeasure against the copy attack that  in order to detect the forgery relies on the high correlation between the PRNU estimate with any of the image residuals used in the estimation. However, to the best of our knowledge, our work constitutes the first attempt at quantifying such leakage by proposing two measures: one based on the mutual information, and another based on the success rate of a membership inference test. Although we do not explicitly try to recover traces of the images used to extract the PRNU, we show that the leakage is large enough to consider the possibility of recovery a serious threat. In this sense, we remark that the images involved in criminal investigations are often of extremely sensitive nature, like in cases involving child abuse and other sexually-oriented crimes, so the mere existence of this leakage calls for the implementation of effective protection mechanisms of the camera fingerprints that ensure that privacy is preserved at all times during investigations. 

Indeed, since law enforcement agencies are accostumed to sharing robust hashes (e.g., those provided by Microsoft's PhotoDNA tool) in order to detect images of child exploitation, there might be as well a predisposition to do so with camera fingerprints. However, as our paper concludes, this should be done only after carefully assessing the risks and considering all the possible remedies, some of which are proposed and discussed in this paper. In particular, we believe that working with encrypted data at all times \cite{pedrouzo}, although yet not entirely practical due to the large amount of computations needed, is the most promising solution.         

The rest of the paper is organized as follows: in Sect.~\ref{sec:prelim} we review the basic principles of PRNU extraction; in Sect.~\ref{sec:measures} we propose two metrics to quantify the leakage; Sect.~\ref{sec:experiments} contains the results of experiments carried on images taken with popular cameras; Sect.~\ref{sec:mitigation} briefly discusses several approaches to mitigate the leakage, and, finally, Sect.~\ref{sec:conclusions} presents our conclusions.

\subsection{Notation} 
Matrices, written in boldface, represent luminance images. All are assumed to be of size $M \times N$. The $(k,l)$th pixel of image $\bt X$ is refered to as $X[k,l]$.
Given two matrices, $\bt X$ and $\bt Y$, its Hadamard product $\bt Z=\bt X \circ \bt Y$ is such that $Z[k,l]=X[k,l]\cdot Y[k,l]$. The Frobenius cross-product of $\bt X$ and $\bt Y$ is defined as $\langle \bt X, \bt Y \rangle_F \doteq \text{tr} \left( \bt X^T \bt Y\right)$, where $\text{tr}(\cdot)$ denotes trace and $T$ transpose. The all-one matrix is denoted by $\bt 1$. Random variables are written in capital letters, e.g., $X$, while realizations are in lowercase, e.g., $x$. Given two random variables $X, Y$, $X \to Y$ means that $X$ converges to $Y$ in probability.       

\section{Preliminaries}
\label{sec:prelim}

In this paper, we will use the prevalent simplified sensor output model presented in \cite{Chen08} in matrix form:
\begin{equation}
    \my \doteq (\bt 1+\mk)\circ\mx+\mn
\end{equation}
where $\my$ is the output of the sensor, $\mk$ is the multiplicative PRNU term, $\mx$ is the noise-free image and $\mn$ collects all the non-multiplicative noise sources. 

This PRNU term can be estimated from a set of $L$ images $\{\my^{(i)}\}_{i = 1}^{L}$ coming from the same sensor. Firstly, the noise-free image $\mx^{(i)}$ is estimated using a denoising filter,\footnote{In the experiments carried out in this paper, we have used the popular wavelet-based denoiser in \cite{Mihcak99}.} and this estimate is used to obtain a residual $\mw^{(i)} \doteq \my^{(i)}-\hat{\mx}^{(i)}$.
Under the assumption of $\mn^{(i)}$ being composed by i.i.d. samples of a Gaussian process, the MLE estimator of $\mk$ reduces to:
\begin{equation}
    \hat{\mk} = \left(\sum_{i = 1}^{L}\mw^{(i)} \circ \hat{\mx}^{(i)}\right) / \ \bt R
\end{equation}
where $\bt R \doteq {\sum_{i = 1}^{L}\hat{\mx}^{(i)} \circ \hat{\mx}^{(i)}}$, and the division is point-wise. Often, the result of this estimation contains non-unique traces left by color interpolation, compression or other systematic errors, that are removed by post-processing (e.g., zero-meaning and Wiener filtering in the full-DFT domain). Ideally, this PRNU will be a zero-mean white Gaussian process with variance $\sigma_k^2$, independent of the location within the matrix. 

Unfortunately, the denoising process will not perform perfectly. In fact, the denoised image can be more accurately modeled as:
\begin{equation}
\hat \mx^{(i)} = \left(\mx^{(i)}-\bt \Delta^{(i)} \right)+\left(\bt 1-\bt \Omega^{(i)} \right)\circ \mk \circ \mx^{(i)}
\end{equation}
where $\bt \Delta^{(i)}$ takes into account the traces of the noise-free image that are left out by the denoising and $\left(\bt 1-\bt \Omega^{(i)} \right)$ models the fraction of the PRNU-dependent component that passes through the denoiser. Then, when subtracted to $\bt Y^{(i)}$ and applied to the estimator, we have:
\begin{equation}
\label{eq:premodel}
\hat{\mk} = \frac{\sum_{i = 1}^{L}\left(\bt \Omega^{(i)}\circ \mk \circ \mx^{(i)}+\bt \Delta^{(i)}+\mn^{(i)}\right) \circ \hat{\mx}^{(i)}}{\bt R}
\end{equation}
Then, it is easy to show that \eqref{eq:premodel} can be expressed as 
\begin{equation}
\label{eq:model} 
    \hat{\mk} = \bt \Omega \circ \mk+\mn_k,
\end{equation}
where $\bt \Omega \doteq \left(\sum_{i = 1}^{L} \bt \Omega^{(i)}\circ \hat{\mx}^{(i)} \circ \mx^{(i)}\right) / \ \bt R$ is a function of the used images and takes into account the amount of PRNU removed in the denoising process, and $\mn_k$ is estimation noise that depends on both $\{\bt \Delta^{(i)}\}_{i = 1}^{L}$ and $\{\bt N^{(i)} \circ \hat \mx^{(i)}\}_{i = 1}^L$,  which in turn convey contextual information about the images. Experiments reported in \cite{gear} show that $\bt N_k$ can be well-modeled by an independent Gaussian process with variance at the $(k,l)$th position denoted by $\gamma^2[k,l]$ .

\begin{figure}[!t]
  \centering
  \subfigure[Original image] {\includegraphics[scale = 0.45]{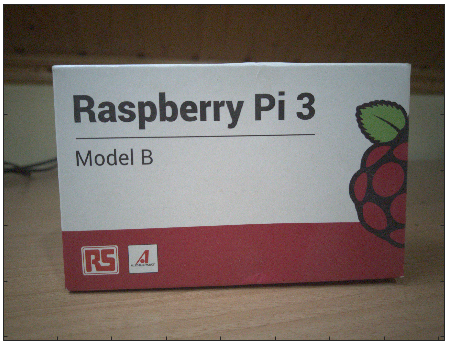}}\quad
  \subfigure[PRNU estimated from the image]{ \includegraphics[scale = 0.44]{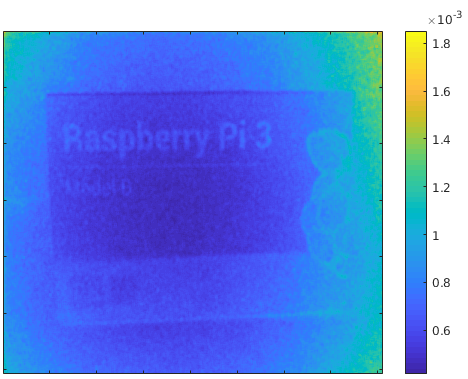}}
  \caption{(a) Sample image containing textual information; (b) PRNU extracted from 24 dark images and the image in (a)}
 \label{fig:rasppi}
\end{figure}

Fig.~\ref{fig:rasppi} illustrates a rather extreme case of leakage in which the PRNU of a Xiaomi MI5S smartphone camera is estimated from 25 images: the one on the upper panel plus 24 additional dark images. As becomes evident, there is a lot of information leaking from the first image into the estimated PRNU. Although by no means this experiment describes a realistic case, it does expose that such alarming leaks may well occur in smaller areas of the image. A more systematic approach to quantifying those leaks is presented in the next section.  

\section{Quantifying the leakage}
\label{sec:measures}
In this section we discuss the two proposed measures to quantify the leakage of the images used for PRNU estimation into the estimate. 

\subsection{Information-theoretic Leakage}

The first measure is based on mutual information of the set of images used for the estimation $\{\bt Y^{(i)}\}_{i=1}^L$ and the estimated PRNU $\hat{\bt K}$, i.e., $I(\{\bt Y^{(i)}\}_{i=1}^L,\hat{\bt K})$. Since $\bt N_k$ is a function of $\{\bt Y^{(i)}\}_{i=1}^L$, we can resort to the data processing inequality to show that $I(\{\bt Y^{(i)}\}_{i=1}^L,\hat{\bt K}) \geq I(\bt N_k,\hat{\bt K})$. The right hand side is considerably simpler to manage and produces an lower bound on the leakage.   

The main difficulty for the calculation of $I(\bt N_k,\hat{\bt K})$ is the lack of a complete statistical characterization for $\bt \Omega$. It has been proven by Ihara \cite{Ihara78} that given a Gaussian process $\bt X$ with covariance $\bt K_x$ and a noise process $\bt Z$ with covariance $\bt K_z$, then the mutual information of $\bt X$ and $\bt X+ \bt Z$ is minimized when $\bt Z$ is Gaussian with covariance $\bt K_z$. Therefore, for a given covariance matrix of $\bt \Omega \circ \mk$, assuming that such process is Gaussian-distributed with the same covariance will produce a lower bound on the mutual information. Now, since $\bt K$ is assumed to be white, its covariance matrix is $\sigma_k^2 \bt I_{MN\times MN}$. Therefore, the covariance of $\bt \Omega \circ \bt K$ will be an ${MN\times MN}$ diagonal matrix with elements $\omega^2[k,l]\sigma_k^2$. Then, the lower-bounding scenario corresponds to $MN\times MN$ parallel channels, in which the 'desired' signal (i.e., $\bt N_k$) is transmitted on each subchannel with power $\gamma^2[l,k]$ and there is an additive Gaussian 'disturbance' (corresponding to $\bt \Omega \circ \bt K$) with power $\omega^2[k,l]\sigma_k^2$. 

Unfortunately, determining $\omega^2[k,l]\sigma_k^2$ turns out to be a difficult problem because even for moderate $L$ the term $\bt N_k$ dominates $\bt \Omega \circ \bt K$ in \eqref{eq:model}. One might think of using flat-field images for this purpose, as in this case the contribution of $\bt N_k$ would be negligible sooner as $L$ increases. However, this path is not advisable because with flat-field images the contribution of $\bt \Omega$ would be lost. Therefore, we must content ourselves with estimating the trace of the covariance matrix of $\bt \Omega \circ \bt K$, given by  $P \doteq \sigma_k^2 \sum_{l,j}\omega^2[l,j]$, and then use it to produce a further lower bound on the mutual information. The value $P$ can be seen as the total disturbance power budget that can be split among the different parallel channels in order to minimize the mutual information. Notice that this represents a worst case because in practice $\sigma_k^2 \omega^2[l,j]$ will deviate at each position $(k,l)$ from such power distribution and the actual leakage will be larger. 

The mutual information in this case can be obtained through the use of Lagrange multipliers, which give the following lower bound \cite{Jorswieck04}:
\begin{equation}
\label{eq:mu1}
  I(\bt N_k,\hat{\bt K}) \geq  \frac{1}{2} \sum_{l,j} \log \left(1+\frac{2}{\sqrt{1+4/(\mu \cdot \gamma^2[l,k])}-1} \right)
\end{equation}
where $\mu$ is the solution to the equation 
\begin{equation}
    \frac{1}{2}\sum_{k,l} \gamma^2[l,k] (\sqrt{1+4/(\mu \cdot \gamma^2[l,k])}-1) = P
\end{equation}

To estimate $P$, we propose to randomly split the set $\{\my^{(i)}\}_{i = 1}^{L}$ into two subsets and estimate $\bt K$ from each. Let $\hat{\bt K}_1$, $\hat{\bt K}_2$ be those estimates. Then, $P$ can be estimated as 
$\hat P= \langle \hat{\bt K}_1, \hat{\bt K}_2 \rangle_F$. A better estimate can be obtained by repeating several times the splitting of $\{\my^{(i)}\}_{i = 1}^{L}$ and averaging the resulting values of $\hat P$. 

We remark here that the leakage that we have quantified through a lower bound corresponds to the complete set of images $\{\my^{(i)}\}_{i = 1}^{L}$ used for estimating $\hat{\bt K}$. This means that we are not quantifying the leakage of a specific image, say, $\my^{(j)}$, $j \in \{1, \cdots, L\}$. Such problem, which is more difficult due to the remaining images acting as a sort of interference, will be the subject of a future work.    


\subsection{Membership inference}
\label{sec:membership_inference}
In the PRNU scenario a membership inference test \cite{inference} is a binary hypothesis test that, given a PRNU estimate, classifies a given image as having been used or not in the estimation. This inference is possible due to the aforementioned leakage: the higher the success rate in the membership inference test, the larger the leakage. It is important to note that the number $L$ of images used in the estimation becomes a key parameter, since as $L$ increases the information provided by the other images will dilute the individual contributions.  


The potential recognition of the images used to estimate the PRNU allows any malicious attacker to obtain information about the input database, which may result in privacy risks in certain scenarios.  As an example, knowing whether certain images were used to compute the PRNU may aid a convicted criminal in identifying the informant who handed them to law enforcement. 

We derive two types of membership detectors: a Neyman-Pearson-based (NP) detector and a normalized-cross-correlation-based (NCC) detector.  Even though the former is expected to perform better due to its statistical properties, along its derivation we will find that it requires information that is not readily available to a potential attacker. Therefore, assuming knowledge of such information leads to a `genie-based' detector which is not practically realizable but is useful as it sets an upper bound on the achievable performance.  In contrast, the NCC detector will behave (slightly) worse but is perfectly implementable.

Let $\my^{(r)}$ be the image whose membership we want to test and which is known to contain the true PRNU $\bt K$. Note that the available observations to implement the test are $\hat{\mx}^{(r)}$, $\mw^{(r)}$ and $\hat{\bt K}$. Then, two hypotheses can be formulated:

\begin{align}
    & \mathcal{H}_0 : \hat{\mk} = \left(\sum_{i = 1}^{L}\mw^{(i)} \circ \hat{\mx}^{(i)}\right) / \ \bt R  \\
    & \mathcal{H}_1 : \hat{\mk} = \bt Q +\left(\sum_{i = 1, i\not= r}^{L}\mw^{(i)} \circ \hat{\mx}^{(i)}\right) / \ \bt R 
\end{align}

where $\bt Q \doteq \left(\mw^{(r)} \circ \hat{\mx}^{(r)}\right) / \bt R$. Matrices $\left(\sum_{i = 1}^{L}\mw^{(i)} \circ \hat{\mx}^{(i)}\right) / \ \bt R$ and  $\left(\sum_{i = 1, i\not= r}^{L}\mw^{(i)} \circ \hat{\mx}^{(i)}\right) / \ \bt R$ can be modeled as having independent zero-mean Gaussian elements with variances at position $(l,j)$ denoted by $\lambda^2_{l, j}$ and $\theta^2_{l, j}$ , respectively.

%

Let $\mathbf{P} \doteq \hat{\mk}-\bt Q$. Then, applying the Neyman-Pearson criterion  \cite{skay}, the following test is obtained: 
\begin{equation}
\label{eq:optimal}
    J\msf{NP} \doteq \sum_{l,j}\left(\log{\left(\frac{\lambda_{l, j}}{\theta_{l, j}}\right)}-\frac{\left({P}[l, j]\right)^2}{2\theta_{l, j}^2} +\frac{\left(\hat{K}[l, j]\right)^2 }{2\lambda_{l, j}^2}\right) > \psi' 
\end{equation}
where $\psi'$ is a threshold selected so that a certain probability of false positive is attained.  
In order to implement the test above, the variances  $\lambda_{l, j}^2$ and $\theta_{l, j}^2$ are needed for all $l,j$. They can be computed as the respective local variances at each position of $\hat{\mk}$ and $\mathbf{P}$. Unfortunately, $\bt P$ is only available through $\bt Q$ that in turn requires knowledge of $\bt R$. Since the latter will be in general unknown to an attacker, the NP detector must be considered only of theoretical interest.  
Notice that when $L \to \infty$, then $\bt P \to \hat \mk$ and $\theta_{l, j}^2 \approx \lambda_{l, j}^2$, for all $l,j$ since the information provided by an individual image is less significant. As a consequence, when $L \to \infty$ the membership test is equivalent to guessing the outcome of (fair) coin tossing.\footnote{This should be reflected in ROC curves as following the `line-of-chance', cf. Sect.~\ref{sec:membership_inference}.}


As a realizable alternative to the NP detector, it is possible to resort to the NCC of $\hat \mk$ and $\bt W^{(r)}$, which has been already employed in camera attribution scenarios \cite{Goljan12}.  This approach relies on the availability of sample estimates of the respective means ($\hat \mu_k$ and $\hat \mu_t$) and variances ($\hat \sigma_k^2$ and $\hat \sigma_t^2$) of   $\hat \mk$ and $\bt W^{(r)}$. The resulting detection statistic becomes
\begin{equation}
\label{eq:ncc}
J\msf{NCC} \doteq \frac{1}{M N-1}\sum_{l, j} \frac{(\hat K[l, j]-\hat \mu_k)}{\hat \sigma_k}\cdot \frac{(W^{(r)}[l, j] -\hat \mu_t)}{\hat \sigma_t}
\end{equation}

\section{Experiments}
\label{sec:experiments}
\subsection{Experimental setup and results}
We have carried out experiments to validate our measures on a database of images taken with several commercially available cameras listed in Table \ref{table:tab1}. The number of images per camera ranges from 122 (Canon1100D\#2) to 316 (Canon1100D\#1). We discuss the results separately for the mutual information and the membership inference test. 

\subsection{Mutual information}
In our first experiment, we have computed the lower bound from \eqref{eq:mu1} (heretofore denoted as Information Leakage Bound, ILB, and measured in bits per pixel, bpp) for two different values of $L$, namely $L=26$ and $L=50$.  The results, shown in Table \ref{table:tab1} correspond to the average ILBs of 10 (resp. 5) runs of the experiment with randomly chosen subsets of size $L=26$  (resp. $L=50$).

\begin{table}[h!]
\centering
\begin{tabular}{| c | c | c| }
\hline
 Camera & ILB ($L=26)$ & ILB ($L=50$) \\ 
 \hline
NikonD60 & 1.6551 & 1.3458 \\
\hline
Canon1100D\#1 & 1.4007 & 1.1037 \\
\hline
Canon1100D\#2 & 1.7100 &  1.4092 \\
\hline
Canon1100D\#3 & 1.5962 & 1.2582 \\
\hline
NikonD3000 & 1.4175 & 1.11467 \\
\hline
NikonD3200 & 1.3827 & 1.0810 \\
\hline
NikonD5100 & 1.9167 & 1.5768 \\ 
\hline
Canon600D &  0.8013  & 0.6791 \\
\hline
NikonD7000 & 1.5246 & 1.2280\\
\hline
\end{tabular}
\caption{Lower bound \eqref{eq:mu1} for different cameras and sizes of estimation sets.}
\label{table:tab1}
\end{table}

As we can see, the leakage (as measured by the ILB), decreases significantly with $L$, as intuition confirms. This can be explained by the fact that the disturbance power budget $P$ stays approximately constant, while the `desired' signal $\bt N_k$ reduces its power with $L$. In fact, notice that, as $L \rightarrow \infty$ the term $\bt N_k$ is expected to go to zero due to the law of large numbers. The relatively small ILBs observed for the Canon 600D camera are conjectured to be due to the images in the respective dataset being very similar to each other. 
                              
In our second experiment, we use images taken with a the camera of a Xiaomi MI5S smartphone to build the following: sets {\tt 50brt} and {\tt 50drk} correspond to $L=50$ images of respectively white and black cardboard, while in sets {\tt 49brt+berry} and {\tt 49drk+berry} one of the images is replaced by the one shown in Fig.~\ref{fig:rasppi}a. The corresponding ILBs are shown in Table \ref{table:tab2}.       

\begin{table}[h!]
\centering
\begin{tabular}{|c|c|c|c|}
\hline
{\tt 50brt} & {\tt 50drk} & {\tt 49brt+berry} & {\tt 49drk+berry}\\
\hline
0.447 & 0.794 & 0.446 & 0.827\\
\hline
\end{tabular}
\caption{Lower bound \eqref{eq:mu1} for flat-field images with and without the image in Fig.~\ref{fig:rasppi}a.}
\label{table:tab2}
\end{table}
As we can see, dark images leak more information. Of course, this leakage does not correspond to perceptually meaningful information, but a look at \eqref{eq:premodel} shows that noise present in any $\bt \Delta^{(i)}$ or $\bt N^{(i)}$ will be boosted by the small denominator in \eqref{eq:premodel}, so that $\bt N_k$ becomes moderately large. For an analogous reason, the opposite effect is observed with bright images, for which the ILB is considerably smaller. Furthermore, while the inclusion of the non-flat image does not increase the information leakage of bright flat-field images, because the former gets diluted in the latter when averaging, this is not the case for dark images: the new image has a considerable impact on $\bt N_k$ and thus contributes to a larger leakage. This is consistent with the empirical observation that it is easier to extract traces from the image in Fig.~\ref{fig:rasppi}a when averaged with dark images (cf. Fig.~\ref{fig:rasppi}b).

\subsection{Membership inference}

\begin{figure}[!t]
    \includegraphics[scale = 0.5]{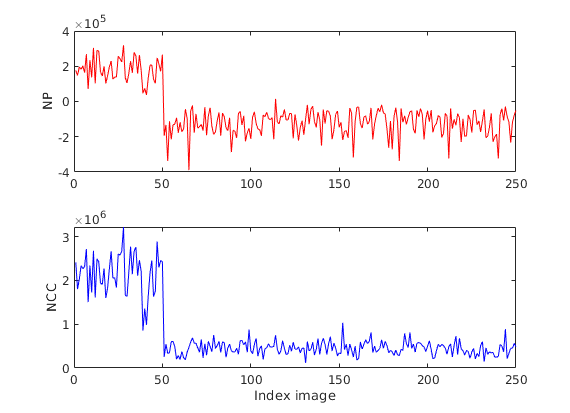}
\caption{Detection statistics for the two detectors on a set of 250 images (Nikon D7000 camera). First 50 images correspond to those used in the PRNU estimation. }
\label{fig:optimal_NCC}
\end{figure}

Aiming at testing the ability and accuracy of both NP and NCC membership inference detectors, several experiments were performed with PRNUs estimated from subsets of 50 images randomly selected from a set of 250 images captured using the NikonD7000 camera. In Fig. \ref{fig:optimal_NCC} the outputs of the NP and NCC detectors are represented for one such subset. The first 50 samples correspond to the membership test statistics for the 50 images used to estimate the PRNU. It is clear from the graphs that both detectors can differentiate which images were used in the estimation as a result of the leakage. 

\begin{figure}[!t]
    \includegraphics[scale = 0.5]{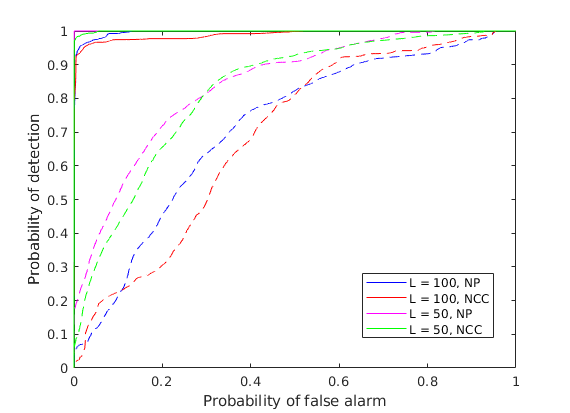}
\caption{Receiver operating characteristic for the NP detector and the NCC, in case L = 100 and L = 50. Solid lines: Nikon D7000; dashed lines: Canon 600D.}\label{fig:roc_curve}
\end{figure}

These results are confirmed by representing the ROC curves for both detectors in Fig. \ref{fig:roc_curve} with $L$ = 100 and $L$ = 50. From this figure the degradation when $L$ increases is again evident. In addition, the NP detector obtains better results, as expected for its being derived from a likelihood ratio. In Fig. \ref{fig:roc_curve} the results for the camera Canon600D are also included. From all our set of cameras, this was the only one in which the membership inference method failed systematically. The reasons why are to be researched yet; there may exist some special property in the PRNU obtained from this camera that hinders the desired information. In any case, these results match those depicted in Tab. \ref{table:tab1}, where the lower bound on the mutual information for this camera is the lowest between all the tested devices. The excellent results obtained with the NikonD7000 are also explainable from the ILBs in the table since this particular model exhibits a high ILB. This confirms the existence of a very close relationship between the membership identification and the lower bound expressed in Eq. \eqref{eq:mu1}, which we intend to explore in the future.

\section{Leakage mitigation}
\label{sec:mitigation}

Given the privacy risks that PRNU leakage entails, it is worth considering potential mitigation strategies. Due to the lack of space, we discuss them briefly here, and leave an in-depth discussion for a future work. We classify countermeasures in three categories: prevention, `deleaking', and privacy preservation. 

Preventive methods aim at conditioning the estimation process so that the resulting PRNU leaks less information. This can be achieved, for instance, by increasing the number of images $L$ whenever possible, maximize the use of flat-field images, or improving denoising algorithms thus reducing $\bt \Delta^{(i)}$ and consequently the leakage, as shown in \eqref{eq:premodel}. 

Deleaking methods consist in modyfing the estimated PRNU in a way that has limited loss (even a gain) in the PRNU detection performance, while decreasing the leakage. Examples of this are PRNU compression methods (e.g.~\cite{Bayram12}), but other possibilities exist, such as whitening the estimated PRNU by normalizing by its local standard deviation at every spatial position. 

Finally, another approach is to limit the exposure of the images and the PRNU in the clear using privacy-preserving techniques. This is possible by carrying out the PRNU estimation with encrypted images (and producing an encrypted PRNU) and detecting the encrypted PRNUs from encrypted query images \cite{pedrouzo}. This way, PRNU detection can be seen as a zero-knowledge proof mechanism. Although this is a very promising approach, substantial work is still needed to reduce the computational complexity of the underlying methods so that they become practical. 

\section{Conclusions}
\label{sec:conclusions}

In this paper, the leakage in the PRNU from the database of images used for its estimation is revealed and lower-bounded using a information-theoretic approach. Experimental results show that this leakage is substantial and thus can entail significant risks to privacy. As a consequence of this leakage, membership identification based on the PRNU becomes possible, achieving high accuracy for both detectors proposed in this paper. 

An open problem which we expect to tackle in the near future is how to use machine learning techniques to reconstruct as reliably as possible the image database from the estimated PRNU. This will illustrate even further the threats to privacy and support the use of leakage mitigation techniques; these will also constitute matter for future work.         


\begin{thebibliography}{1}


 
 \bibitem{Chen08} M. Chen, J. Fridrich, M. Goljan, and J. Lukas, “Determining image origin and integrity using sensor noise,” IEEE Trans. Inf. Forensics Security, vol. 3, no. 1, pp. 74–90, Mar. 2008.
 
 \bibitem{KH17}
P. Korus and J. Huang, “Multi-scale analysis strategies in PRNU-based tampering localization,” IEEE Trans. Inf. Forensics Security, vol. 12, no. 4, pp. 809–824, Apr. 2017.


\bibitem{Cozzolino19}
 D. Cozzolino and L. Verdoliva, "Noiseprint: a CNN-based camera model fingerprint," in IEEE Trans. Inf. Forensics and Security, to be published, 2019.
 
 \bibitem{triangle_test}
 M. Goljan, J. Fridrich, M. Chen, "Defending Against Fingerprint-Copy Attack in Sensor-Based Camera Identification," IEEE Transactions on Information Forensics and Security, vol. 6. pp. 227 - 236, 2011.
 

\bibitem{pedrouzo}
A. Pedrouzo-Ulloa, M. Masciopinto, J.R. Troncoso-Pastoriza, F. P\'erez-Gonz\'alez, “Camera Attribution Forensic Analyzer in the Encrypted Domain,'' Proc. IEEE International Workshop on Information Forensics and Security (WIFS), Hong-Kong, 2018, pp. 1-7. 
 
 \bibitem{gear}
M. Masciopinto and F. Pérez-González, "Putting the PRNU Model in Reverse Gear: Findings with Synthetic Signals," 26th European Signal Processing Conference (EUSIPCO), Rome, pp. 1352-1356, 2018. 



\bibitem{skay}
S.M. Kay. "Fundamentals of statistical signal processing: detection theory." Prentice-Hall, Inc., USA, 1998. 

\bibitem{inference}
R. Shokri, M. Stronati, C. Song and V. Shmatikov, "Membership Inference Attacks Against Machine Learning Models," 2017 IEEE Symposium on Security and Privacy (SP), San Jose, CA, 2017, pp. 3-18.


\bibitem{Jorswieck04}
E. Jorswieck and H. Boche, "Performance Analysis of Capacity of MIMO Systems under Multiuser Interference Based on Worst-Case Noise Behavior", EURASIP Journal on Wireless Communications and Networking, vol. 2, pp. 273-285, 2004. 

\bibitem{Ihara78}
S. Ihara,"On the Capacity of Channels with Additive Non-Gaussian Noise", Information and Control, vol. 37, no. 1, pp. 34-39, 1978. 

\bibitem{Mihcak99} M.K. Mihcak,   I. Kozintsev,     and   K. Ramchandran,   “Spatially   Adaptive Statistical Modeling of Wavelet Image Coefficients and its Application to Denoising.” Proc. IEEE Int. Conf. Acoustics, Speech, and  Signal  Processing,  Phoenix,  AZ,  vol.  6,  pp.  3253–3256,  March  1999 

\bibitem{Goljan12} M. Goljan, J. Fridrich,  “Sensor-Fingerprint Based Identification of Images Corrected for Lens Distortion.” Proc.  SPIE - The International Society for Optical Engineering vol. 8303, February 2012. 

\bibitem{Bayram12} S. Bayram, H. T. Sencar, and N. Memon, “Efficient sensor fingerprint
matching through fingerprint binarization,” IEEE Trans. Inf. Forensics Security, vol. 7, no. 4, pp. 1404–1413, Aug. 2012.



\end{thebibliography}

\end{document}